\documentclass[letterpaper]{article}

\usepackage{aaai}
\usepackage{times}
\usepackage{helvet}
\usepackage{courier}
\usepackage{graphicx}
\usepackage{amsmath}
\usepackage{epsfig}
\usepackage{subfigure}
\usepackage{epstopdf}
\usepackage{paralist}
\usepackage{bm}

\begin{document}
%
\title{Sequential Click Prediction for Sponsored Search \\ with Recurrent Neural Networks}

\author{
Yuyu Zhang\thanks{This work was performed when the first three authors were visiting Microsoft Research Asia.}\\
Institute of Computing Technology\\
Chinese Academy of Sciences\\
zhangyuyu@ict.ac.cn
\And
Hanjun Dai$^*$\\
Fudan University\\
daihanjun@gmail.com
\And
Chang Xu$^*$\\
Nankai University\\
changxu0731@gmail.com
\And
Jun Feng\\
Tsinghua University\\
feng-j13@mails.tsinghua.edu.cn
\AND
Taifeng Wang\\
Microsoft Research\\
taifengw@microsoft.com
\And
Jiang Bian\\
Microsoft Research\\
jibian@microsoft.com
\And
Bin Wang\\
Institute of Computing Technology\\
Chinese Academy of Sciences\\
wangbin@ict.ac.cn
\And
Tie-Yan Liu\\
Microsoft Research\\
tyliu@microsoft.com
}

\maketitle

\begin{abstract}
    \begin{quote}
    Click prediction is one of the fundamental problems in sponsored search. Most of existing studies took advantage of machine learning approaches to predict ad click for each event of ad view independently. However, as observed in the real-world sponsored search system, user's behaviors on ads yield high dependency on how the user behaved along with the past time, especially in terms of what queries she submitted, what ads she clicked or ignored, and how long she spent on the landing pages of clicked ads, etc. Inspired by these observations, we introduce a novel framework based on Recurrent Neural Networks (RNN). Compared to traditional methods, this framework directly models the dependency on user's sequential behaviors into the click prediction process through the recurrent structure in RNN. Large scale evaluations on the click-through logs from a commercial search engine demonstrate that our approach can significantly improve the click prediction accuracy, compared to sequence-independent approaches.
    \end{quote}
\end{abstract}

\section{Introduction}\label{sec:introduction}

Sponsored search has been a major business model for modern commercial Web search engines. Along with organic search results, it presents to users with sponsored search results, i.e., advertisements (ads) targeting to the search query. Sponsored search accounts for the overwhelming majority of income for three major search engines: Google, Yahoo and Bing. Even in the US search market alone, it generates over 20 billion dollars per year, the amount of which still keeps rising\footnote{Source: eMarketer, June 2013.}. According to the common cost-per-click (CPC) model for sponsored search, advertisers are only charged once their advertisements are clicked by users. In this mechanism, to maximize the revenue for search engine and maintain a desirable user experience, it is crucial for search engines to estimate the click-through rate (CTR) of ads.

Recently, click prediction has received much attention from both industry and academia~\cite{fain2006sponsored,jansen2008sponsored}. State-of-the-art sponsored search systems typically employ machine learning approaches to predict the click probability by using the feature extracted based on: 1) historical CTR for ad impressions~\footnote{We refer to a certain ad shown to a particular user in a specific search result page as an ad impression.} with respect to different elements, e.g., CTR of query, ad, user, and their combinations~\cite{graepel2010web,richardson2007predicting}; 2) semantic relevance between query and ad~\cite{radlinski2008optimizing,shaparenko2009data,hillard2011sum}.

However, most of previous works take single ad impression as the input instance to predict the click probability, without considering dependency between different ad impressions. Some recent research work~\cite{xu2010temporal,xiong2012relational} pointed out that they could achieve more accurate click prediction by modeling spatial relationship between ad slots in the same query session. Inspired by them, we conduct further data analysis to study other types of dependency between user's behaviors in sponsored search, through which we find that user's behaviors also yield explicit temporal dependency. For example, if a user clicks an ad, comes to the ad landing page, but closes it very quickly, the click probability of her next view of this ad will become fairly low; moreover, if a user has previously submitted a query on booking flight, he/she will be with higher probability to click the ads under flight booking. These findings motivate us to advance the state-of-the-art of click prediction by modeling the important temporal dependency into the click prediction process. Although some kinds of dependency can be modeled as features, it's still hard to identify all of them explicitly. Thus, it is necessary to empower the model with the ability to extract and leverage various kinds of dependency automatically.

In real-world sponsored search system, the event of any ad impression, click, and corresponding context information (e.g. user query, ad text, click dwell time, etc.) is recorded with the time stamp in search logs. Thus, it is natural to employ time series analysis methods to model sequential dependency between user's behaviors. Previous studies on time series analysis~\cite{kirchgassner2012introduction,box2013time} usually focused on modeling trends or periodic patterns in data series. However, the sequential dependency between ad impressions is so complex and dynamic that time series analysis approaches is not capable enough to model it effectively. On the other hand, a few most recent studies leverage Recurrent Neural Networks (RNN) to model the temporal dependency in data. For example, RNN language model~\cite{mikolov2010recurrent,mikolov2011extensions,mikolov2011rnnlm} successfully leverages long-span sequential information among the massive language corpus, which results in better performance than traditional neural networks language model~\cite{bengio2006neural}. Moreover, RNN based handwriting recognition~\cite{graves2009novel}, speech recognition~\cite{kombrink2011recurrent}, and machine translation~\cite{auli2013joint} systems have also led to much improvement in the corresponding tasks. Compared to traditional feedforward neural networks, RNN has demonstrated its strong capability to exploit dependencies in the sequence due to its specific recurrent network structure.

In this work, we propose to leverage RNN to model sequential dependency into predicting ad click probability. We consider each user's ad browsing history as one sequence which yields the intrinsic internal dependency. In the training process of RNN model, features of each ad impression will be feedforwarded into the hidden layer, together with previously accumulated hidden state. In this way, the dependency among impressions will be embedded into the recurrent network structure. Our experiments on the large scale data from a commercial search engine reveal that, such RNN structure can give rise to a significant improvement on the click prediction accuracy compared with the state-of-the-art dependency-free models such as Neural Networks and Logistic Regression.

The main contributions of this paper are in three folds:
\begin{itemize}
  \item We investigate the sequential dependency among particular user's ad impressions, and identify several important sequential dependency relationships.
  \item We use Recurrent Neural Networks to model user's click sequence, and successfully incorporate sequential dependency into enhancing the accuracy of click prediction.
  \item We conduct large scale experiments to validate the RNN model's effectiveness for modeling sequential data in sponsored search.
\end{itemize}

In the following parts of this paper, we will first present our data analysis results to verify the potential dependency which might affects click prediction. Then, we propose our RNN model for the task of sequence-based click prediction. After that, we describe experimental settings followed by the experimental results and further performance study. At last, we summarize this paper and discuss some future work.

\section{Data Analysis on Sequential Dependency}\label{sec:data_analysis}

To gain more understanding on why sequential information is important in click prediction, in this section, we will discuss the effects of sequential dependency from multiple perspectives. We collect data for analysis from the logs of a commercial sponsored search system.

\begin{figure}
\centering
\includegraphics[width=.42\textwidth]{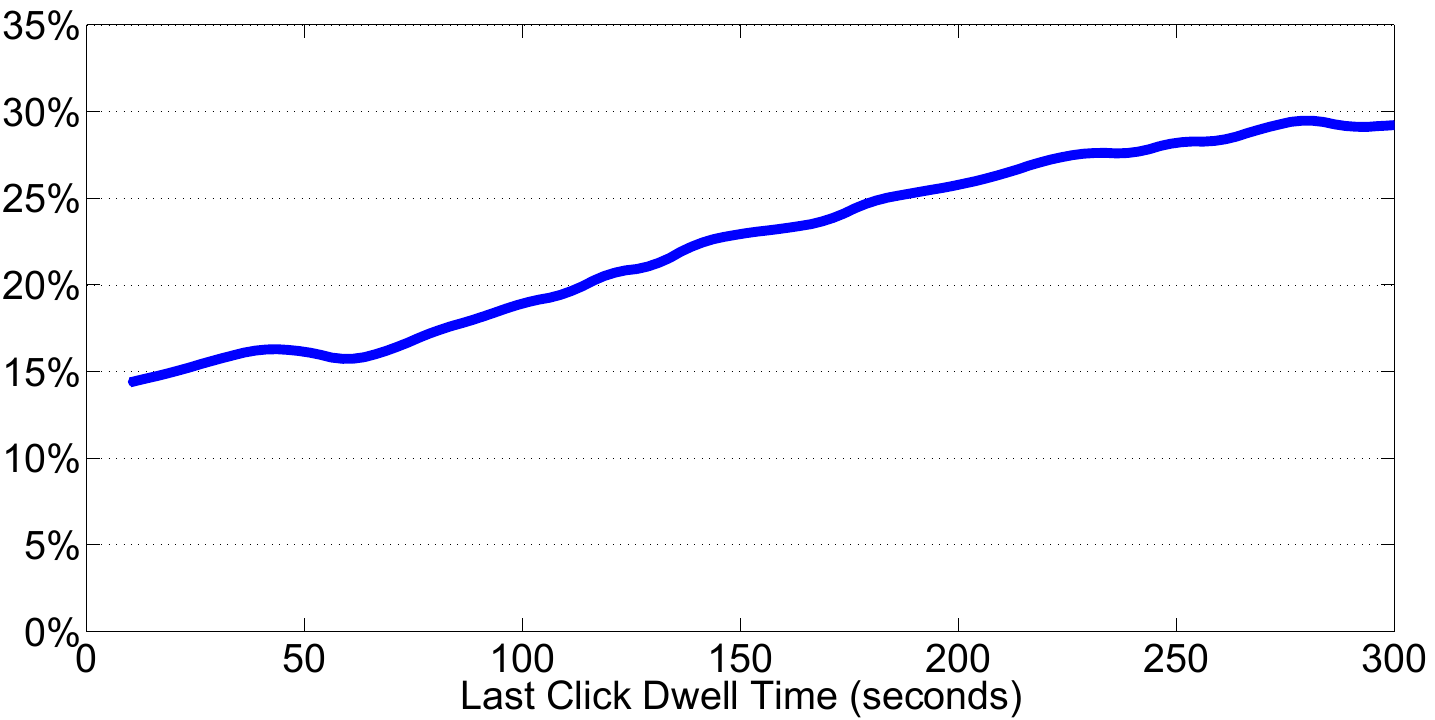}
\caption{Correlation between last click dwell time and current click-through rate.}
\label{fig:dwelltime_ctr}
\end{figure}

Once a user clicks an ad, she will enter into the corresponding ad landing page and stay for a certain period of time, which is referred to as the click dwell time. Generally, longer dwell time implies better user experience. For a particular user, all the ad impressions can be organized as an ordered sequence along with the time. To explore the sequential effect of click dwell time, we first pick up all the ``first clicks'' in each user's sequence, and track whether each ad will be clicked again in its consecutively next impression. The correlation between the previous click dwell time and the current click-through rate\footnote{In this paper, we present relative click-through rate to preserve the proprietary information.} is shown in Figure~\ref{fig:dwelltime_ctr}. From this figure, it is clear to observe an obvious positive correlation, i.e., the longer a user stays on an ad's landing page, the more likely she will click this ad right at the next time.

\begin{figure}
\centering
\includegraphics[width=.415\textwidth]{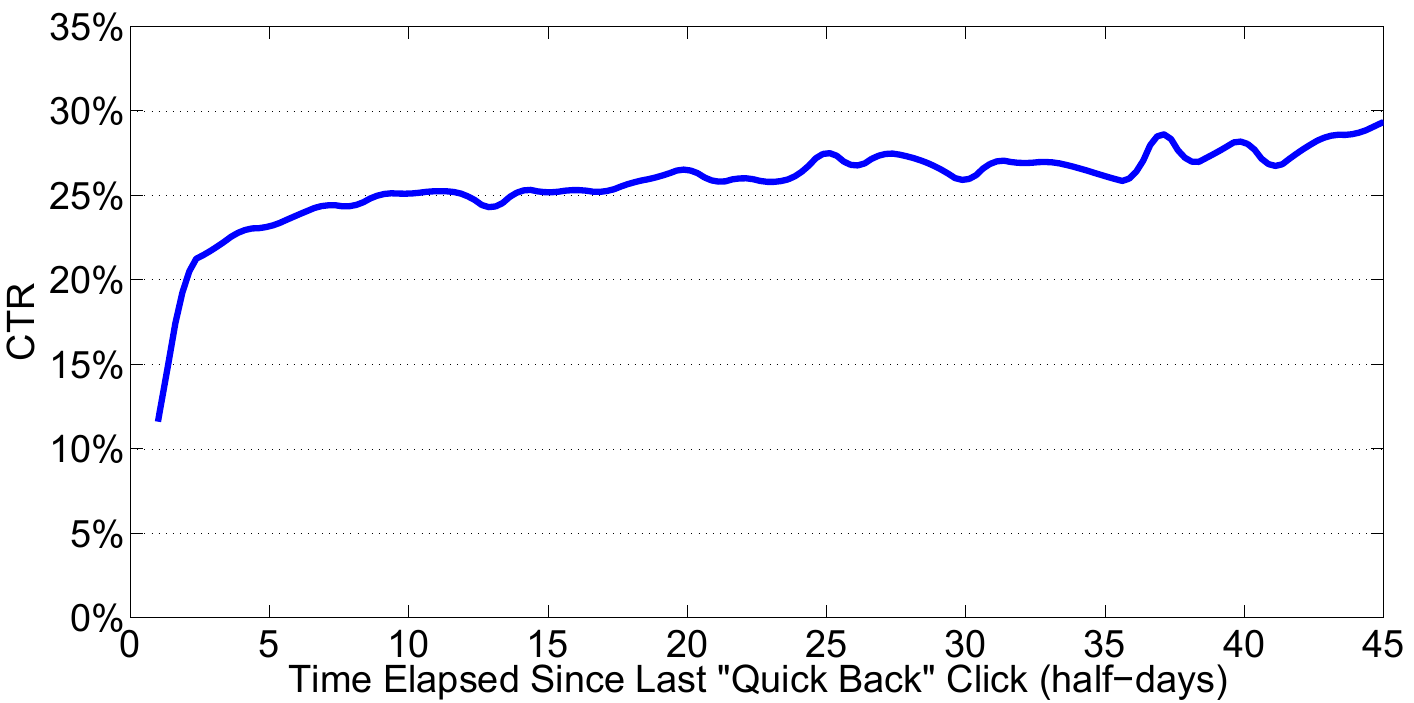}
\caption{Click-through rate right after a ``quick back'' click on the same ad.}
\label{fig:dwelltime_quickback}
\end{figure}

When the click dwell time is less than 20 seconds, we call such ad click as a ``quick back'' one. From the data analysis above, we can observe that ``quick back'' can give rise to rather lower click-through rate in the next impression, which indicates that users tend to avoid clicking the ad once they had an unsatisfied experience. However, it is not clear how users behave if they experienced a ``quick back'' click long time ago (e.g., half a month). To explore this, we collect all the ``quick back'' clicks and calculate the click-through rate after different time intervals, i.e., the time elapsed since the ``quick back'' click, respectively. Figure~\ref{fig:dwelltime_quickback} shows the result, where the time interval is binning by half-days. As conjectured, along with increasing elapsed time, the overall click-through rate grows significantly and then stay steadily in a certain level, which implies that users tend to gradually \mbox{\emph{forget}} the unsatisfied experience with the time passing.

\begin{figure}
\centering
\includegraphics[width=.402\textwidth]{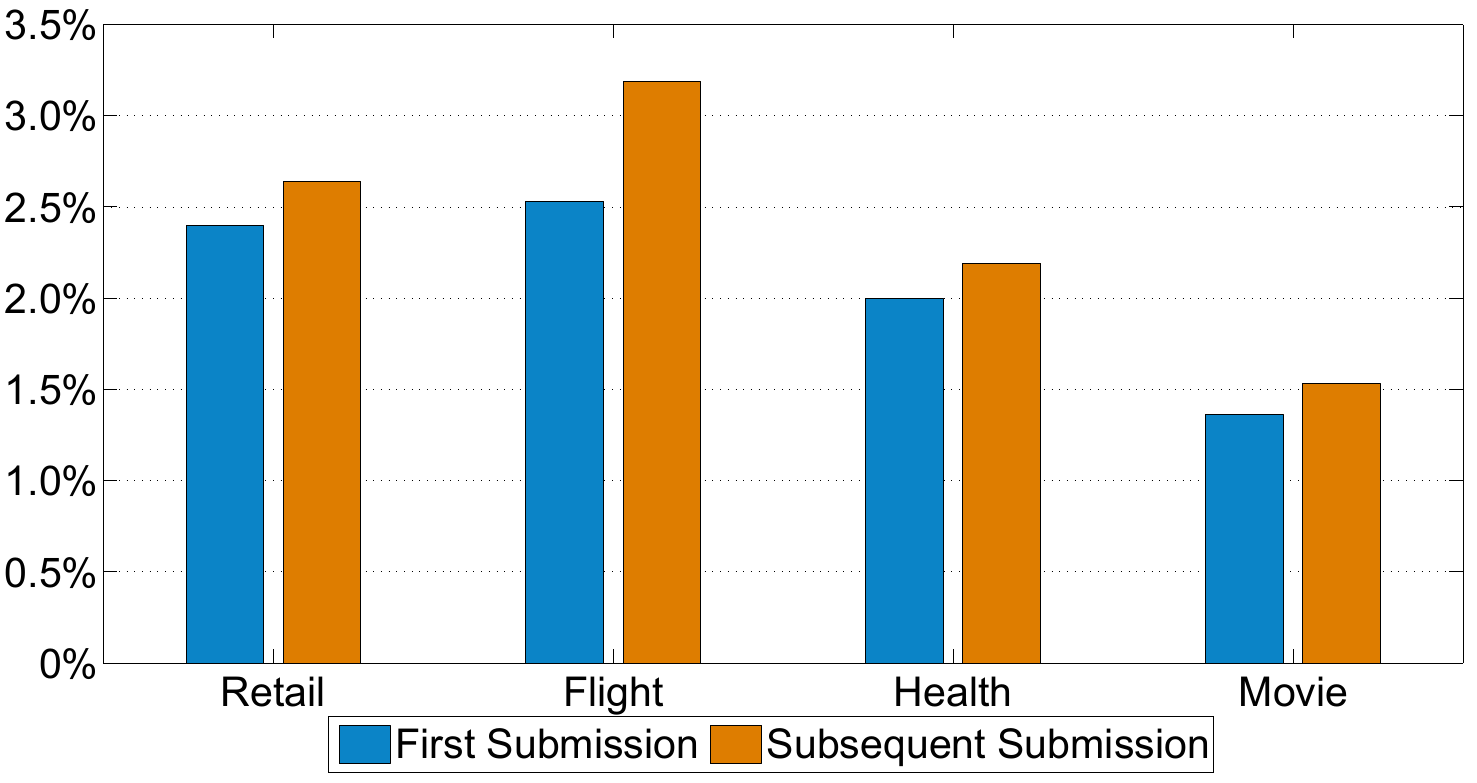}
\caption{Click-through rate on users' first and subsequent submissions of a certain type of query.}
\label{fig:querytype_ctr}
\end{figure}

Besides studies on the sequential effect of dwell time, we further analyze such effects from the aspect of query. In sponsored search systems, ads are automatically selected by matching with the user submitted query. Queries also form a time ordered sequence, binding on that of ad impressions. After categorizing the queries into topics based on our proprietary query taxonomy, we calculate users' click-through rate when they submit each query topic for the first time, and compare with their future CTR on subsequent queries of the same query topic. Analysis results, as shown in Figure~\ref{fig:querytype_ctr}, illustrate that if a user has submitted a query belonging to a certain query topic, he/she will become more likely to click the ads under the same topic.

All the analysis results above indicate that user's previous behaviors in sponsored search may cause strong but quite dynamic impact her subsequent behaviors. As long as we can identify such sequential dependency between user behaviors, we can design features accordingly to enhance the click prediction. However, since a big challenge to enumerate such dependency in the data manually, it is necessary to let the model have the ability to learn such kind of dependency by itself. To this end, we propose to leverage a widely used framework, Recurrent Neural Network, as our model, as it naturally embeds dependencies in the sequence.

\section{The Proposed Framework}\label{sec:the_proposed_framework}

    \subsection{Model}

    \begin{figure}
    \centering
    \includegraphics[width=.47\textwidth]{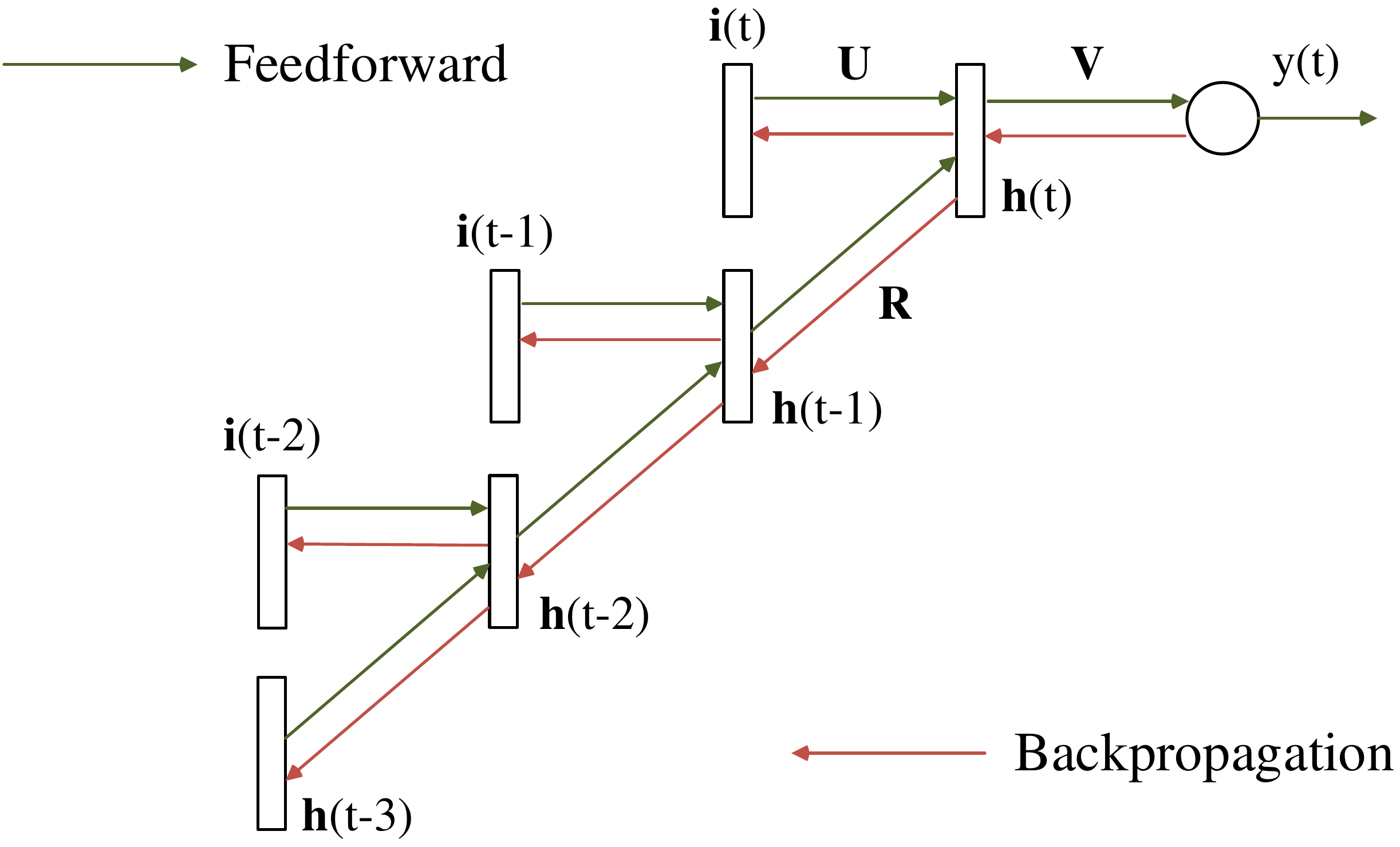}
    \caption{RNN training process with BPTT algorithm. Unfolding step is set to 3 in this figure.}
    \label{fig:bptt_train}
    \end{figure}

    The architecture of a recurrent neural network is shown in Figure~\ref{fig:bptt_train}. It consists of an input layer $i$, an output unit, a hidden layer $h$, as well as inner weight matrices. Here we use $t \in \mathbf{N} $ to represent the time stamp. For example, $\bm{h}(t)$ denotes the hidden state in time $t$. Specifically, the recurrent connections $\bm{R}$ between $\bm{h}(t-1)$ and $\bm{h}(t)$ can propagate sequential signals. The input layer consists of a vector $\bm{i}(t)$ that represents the features of current user behaviors, and the vector $\bm{h}(t-1)$ represents the values in the hidden layer computed from the previous step.

    The activation values of the hidden and output layers are computed as
    \[
    	\bm{h}(t) = f(\bm{i}(t)\bm{U}^T + \bm{h}(t-1)\bm{R}^T),
    \]
    \[
      	y(t) = \sigma(\bm{h}(t)\bm{V}^T),
    \]
    where $f(z) = \frac{1-e^{-2z}}{1+e^{-2z}}$ is the $tanh$ function we use for non-linear activation, and $\sigma(z) = \frac{1}{1+e^{-z}}$ is the sigmoid function for predicting the click probability.

    The hidden layer can be considered as an internal memory which records dynamic sequential states. The recurrent structure is able to capture a long-span history context of user behaviors. This makes RNN applicable to the tasks related to sequential prediction.

    In our framework, $\bm{i}(t)$ represents the features correlated to user's current behavior and $\bm{h}(t)$ represents sequential information of user's previous behaviors. Thus, our prediction depends on not only the current input features, but also the sequential historical information.

    \subsection{Feature Construction}

    In our study, we take ad impressions as instances for both model training and testing. Based on the rich impression-centric information, we construct the input features that can carry crucial information to achieve accurate CTR prediction for a given impression. All these features can be grouped into several general categories: 1) \emph{Ad features} consist of information about ad ID, ad display position, and ad text relevance with query. 2) \emph{User features} include user ID and query (submitted by user) related semantic information. 3) \emph{Sequential features} include time interval since the last impression, dwell time on the landing page of the last click event, and whether the current impression is the head of sequence, i.e., whether it is the first impression for this user.

    With all these diverse types of features, each ad impression can be described in a large and complex feature space. In this paper, all of the click prediction models will follow this input feature setting, so that we can fairly compare their capabilities of predicting whether an impression will be clicked by user.

    \subsection{Data Organization}

    To effectively harness the sequential information for modeling temporal dependencies, the training process requires large quantity of data. Luckily, the data in computational advertising is big enough to make RNN tractable. To obtain the data for sequential prediction, we re-organize the input features along with the user dimension (i.e., reordered each user's historical behaviors according to the timeline). In particular, for the ads in the same search session, we rank them by their natural display orders in the mainline and then sidebar.

    \subsection{Learning}

        \subsubsection{Loss Function}

        In our work, the loss function is defined as an averaged cross entropy, which aims at maximizing the likelihood of correct prediction,
        \[
        L = \frac{1}{M}\sum_{i=1}^M (-y_i log(p_i) - (1-y_i) log(1-p_i)),
        \]
        where M is the number of training samples. The $ith$ sample is labeled with $y_i \in \{0, 1\}$ and $p_i$ is the predicted click probability of the given ad impression.

        \subsubsection{Learning Algorithm (BPTT)}

        RNN can be trained in the same way as normal feedforward network using backpropagation algorithm. In this way, basically, the state of the hidden layer from previous time step is simply regarded as an additional input. With only one hidden layer, the network tries to optimize prediction of the next sample given the previous sample and previous hidden state. However, no effort is directly devoted towards longer context information, which may hurt the performance of RNN.

        A simple extension of the training algorithm is to unfold the network and backpropagate errors even further. This is called Back Propagation Through Time (BPTT) algorithm. BPTT was proposed in~\cite{rumelhart2002learning}, and has been used in the practical application of RNN language model~\cite{mikolov2012statistical}.

        We illustrate the overall training pipeline that applies BPTT to the RNN based click prediction models in Figure~\ref{fig:bptt_train}. Such unfolded RNN can be viewed as a deep neural network with $T$ hidden layers where the recurrent weight matrices are shared and identical. In this approach, the hidden layer can actually exploit the information of the most recent inputs and put more importance to the latest input, which is essentially coherent with the the sequential dependency. In the following part, we use $T$ to denote the number of unfolding steps in the BPTT algorithm.

        The network is trained by Stochastic Gradient Descent (SGD). The gradient of the output layer is computed as
        \[
                e_o(t) = y(t) - l(t),
        \]
        where $y(t)$ is the predicted click probability, and $l(t)$ is the binary true label according to the ad is clicked or not. The weights $\bm{V}$ between the hidden layer $\bm{h}(t)$ and output unit $y(t)$ are updated as
        \[
                \bm{V}(t+1) = \bm{V}(t) - \alpha \times e_o(t) \times \bm{h}(t),
        \]
        where $\alpha$ is the learning rate. Then, gradients of errors are propagated from the output layer to the hidden layer as
        \[
                \bm{e_h}(t) = e_o(t)\bm{V} * (\bm{\vec{1}} - \bm{h}(t)*\bm{h}(t)),
        \]
        where $*$ represents the element-wise product, and $\bm{\vec{1}}$ is a vector with all elements equal to one.

        Errors are also recursively propagated from the hidden layer $\bm{h}(t-\tau)$ to the hidden layer from previous step $\bm{h}(t-\tau-1)$, that is
        \[
                \bm{e_h}(t-\tau-1) = \bm{e_h}(t-\tau)\bm{R} * (\vec{\mathbf{1}} - \bm{h}(t-\tau-1) * \bm{h}(t-\tau-1)),
        \]
        where $\tau \in [0, T)$. The weight matrix $\bm{U}$ and the recurrent weights $\bm{R}$ are then updated as
        \[
                \bm{U}(t+1) = \bm{U}(t) - \alpha \left[ \sum_{z=0}^{T-1}{\bm{e_h}(t-z)^T\bm{i}(t-z)} \right],
        \]
        \[
                \bm{R}(t+1) = \bm{R}(t) - \alpha \left[ \sum_{z=0}^{T-1}{\bm{e_h}(t-z)^T\bm{h}(t-z-1)} \right].
        \]

        Note that, in our practical experiments, we add the bias terms and L2 penalty of weights to the model, and the gradients can still be computed easily based on a slight modification to the equations above.

    \subsection{Inference}

    \begin{figure}
    \centering
    \includegraphics[width=.47\textwidth]{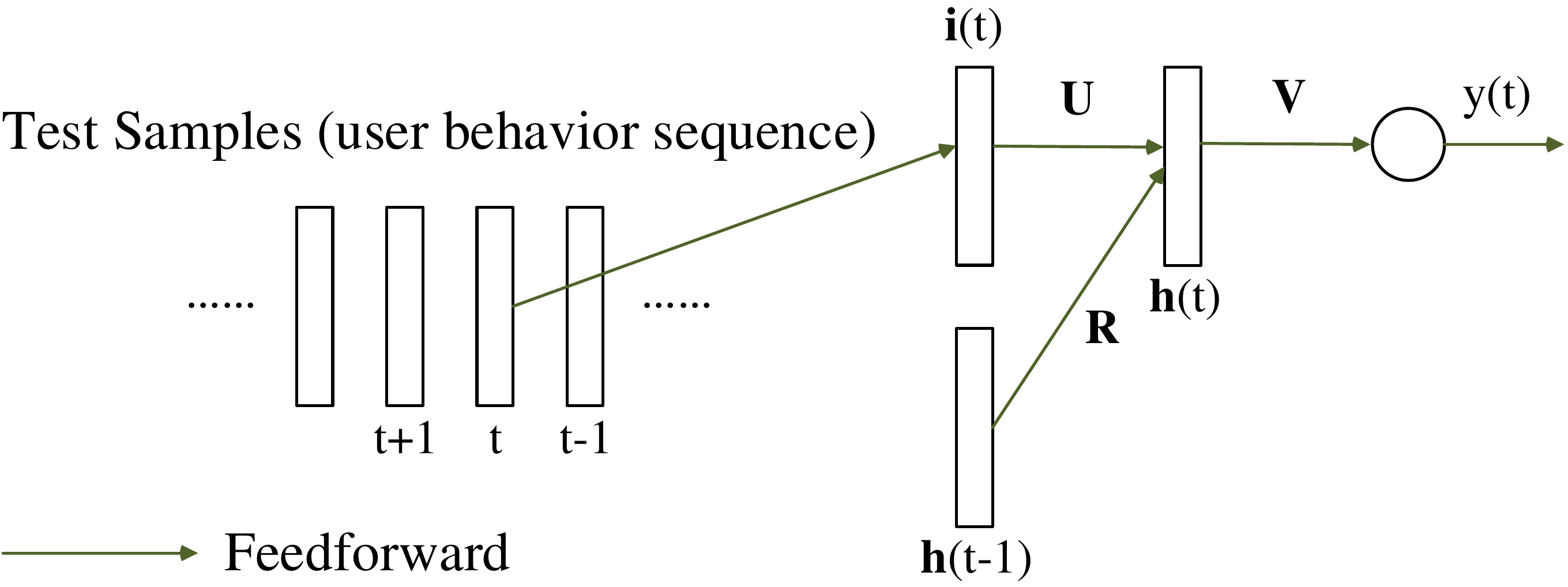}
    \caption{RNN testing process with sequential input samples. The hidden state of previous test sample will be used as input, together with the current sample features.}
    \label{fig:bptt_test}
    \end{figure}

    In contrast to traditional neural networks, RNN has a recurrent layer to store the previous hidden state. In the inference phase, we still need to store the hidden state of previous test sample, and feedforward it with the recurrent weights $\bm{R}$.

    Figure~\ref{fig:bptt_test} illustrates the testing process. The test data is also organized as ordered user behavior sequences. We feedforward current sample features, together with the hidden state of previous sample to get the current hidden state. Then we make the prediction and replace the stored hidden state with current values. Here we only record the hidden state of the last test sample, no matter how many unfolding steps there are in the BPTT training process.

\section{Experiments}\label{sec:experiments}

This section first describes the settings of our experiments, and then reports the experimental results.

    \subsection{Data Setting}

    To validate whether the RNN model we proposed can really help enhance the click prediction accuracy, we conduct a series of experiments based on the click-through logs of a commercial search engine. In particular, we collect half a month logs from November 9th to November 22nd in 2013 as our experimental dataset. And, we randomly sample a set of search engine users (fully anonymized) and collect their corresponding events from the original whole traffic. Finally, we collect over 7 million ad impressions in this period of time. After that, we use the first week's data to train click prediction models, and apply those models to the second week's data for testing. Detailed statistics of the dataset can be found in Table~\ref{table:data_stats}.

    \begin{table}
    \centering
    \caption{Statistics of the dataset for training and testing click prediction models.}
    \label{table:data_stats}
    \vspace{5pt}
    \begin{tabular}{|l|l|l|l|}
      \hline
      & Ad Impressions & Ads & Users \\
      \hline
      Training & 3,740,980 & 1,363,687 & 235,215 \\
      \hline
      Testing & 3,741,500 & 1,379,581 & 235,215 \\
      \hline
    \end{tabular}
    \end{table}

    \subsection{Evaluation Metrics}

    In our work, there are multiple models to be applied to predict the click probability for ad impressions in the testing dataset. We use recorded user actions, i.e., click or non-click, in logs as the true labels. To measure the overall performance of each model, we follow the common practice in previous click prediction research in sponsored search and employ Area Under ROC Curve (\textbf{AUC}) and Relative Information Gain (\textbf{RIG}) as the evaluation metrics~\cite{graepel2010web,xiong2012relational,wang2013exploring}.

    \subsection{Compared Methods}

    In order to investigate the model effectiveness, we compare the performance of our RNN model with other classical click prediction models, including Logistic Regression (LR) and Neural Networks (NN), with identical feature set as described aforementioned. We set LR and NN as baseline models due to the following reasons: 1) Quite a few previous studies~\cite{richardson2007predicting,mcMahan2013ad,wang2013exploring} have demonstrated that they are state-of-the-art models for click prediction in sponsored search. 2) LR and NN models ignore the sequential dependency among the data, while our RNN based framework is able to model such information. Through the comparison with them, we will see whether RNN can successfully leverage dependencies in the data sequence to help improve the accuracy of click prediction.

	\subsection{Experimental Results}

    \subsubsection{Overall Performance}
	
	\begin{table}
    \centering
    \caption{Overall performance of different models in terms of AUC and RIG.}
    \label{table:overall_performance}
    \vspace{5pt}
    \begin{tabular}{|l|l|l|}
      \hline
      Model & AUC & RIG \\
      \hline\hline
      LR & 87.48\% & 22.30\% \\
      \hline
      NN & 88.51\% & 23.76\% \\
      \hline
      RNN & \textbf{88.94\%} & \textbf{26.16\%} \\
      \hline
    \end{tabular}
    \end{table}
	
    For fair model comparison, we carefully select the parameters of each model with cross validation, and ensure every model achieve its best performance respectively. To be more specific, parameters for grid search include: the coefficient of L2 penalty, the number of training epochs, the hidden layer size for RNN and NN models, and the number of unfolding steps for RNN. Finally, we get the best settings of parameters as follows: the coefficient of L2 penalty is $1e-6$, the number of training epochs is 3, the hidden layer size is 13, and the number of unfolding steps should be 3 (more details will be provided later).

    Table~\ref{table:overall_performance} reports the overall AUC and RIG of all three methods on test dataset. It demonstrates that our proposed RNN model can significantly improve the accuracy of click prediction, compared with baseline approaches. In particular, in terms of \textbf{RIG}, there is about \textbf{17.3\%} relative improvement over LR, and about \textbf{10\%} relative improvement over NN. As for the metric of \textbf{AUC}, we can find there is about \textbf{1.7\%} relative gain over LR, and about \textbf{0.5\%} relative gain over NN. In real sponsored search system, such improvement in click prediction accuracy will lead to a significant revenue increment.

    The overall performance above shows the effectiveness of our RNN model, which clearly transcends sequence independent models. Next, we will conduct detailed analysis on how the sequential information help to get more accurate click prediction.
	
    \subsubsection{Performance on Specific Ad Positions}

    It is well-known that the click-through rate on different ad positions varies a lot, which is often referred to as the position bias. To further check the performance of models within specific positions, in our experiments, we separately analyze the performance of RNN model and two baseline algorithms on different ad positions: top first, mainline and sidebar.
	
\begin{figure}
		\centering
		\subfigure[AUC on specific ad positions \label{fig:auc_position}]
		{
			\includegraphics[width=.403\textwidth]{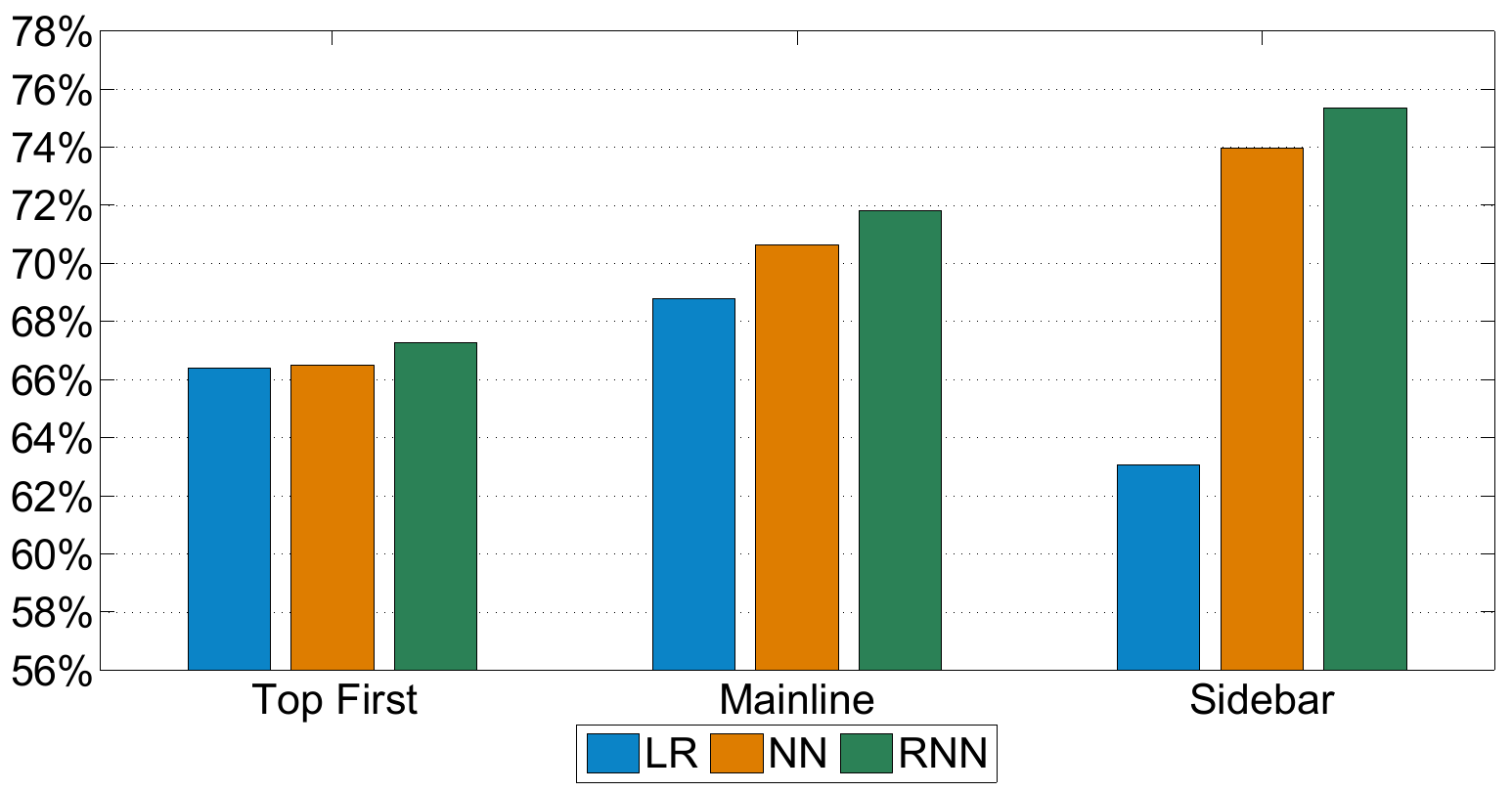}
		}
		\subfigure[RIG on specific ad positions \label{fig:rig_position}]
		{
			\includegraphics[width=.4\textwidth]{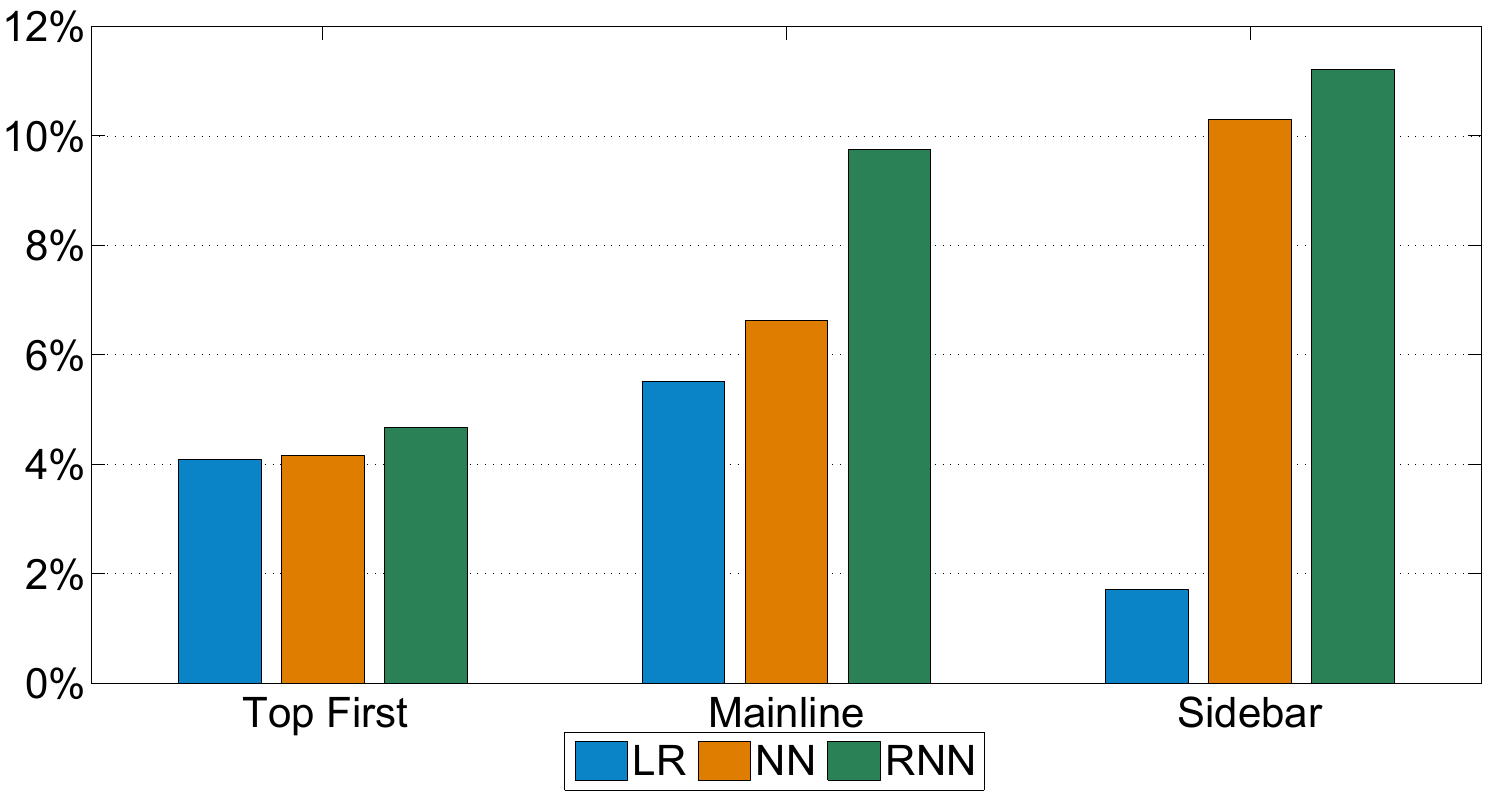}
		}
		\caption{Performance on specific ad positions.}
		\label{fig:performance_position}
    \end{figure}

    Figure~\ref{fig:performance_position} shows the evaluation results on different positions. In Figure~\ref{fig:auc_position}, RNN outperforms NN and LR measured by AUC on all positions. In Figure~\ref{fig:rig_position}, in terms of RIG, RNN achieves impressive relative gain over NN and LR, especially on mainline positions, where RNN beats NN by \textbf{3.12\%}. According to our statistics on daily traffic data, most of revenue comes from the mainline ad clicks, where RNN can achieve significantly better performance. While for sidebar positions, the ads shown there are easily to be ignored by users, so that the clicks or positive instances are very rare. This may drastically hurt the performance of LR model. Nevertheless, the RNN model still performs the best, which indicates that even in rare cases, sequential information can still help.
	
    \subsubsection{Effect of Recurrent State for Inference}

    We have shown the inference method of RNN in Figure~\ref{fig:bptt_test}. To further verify the importance of utilizing historical sequences in the inference phase, we remove the recurrent part of the RNN model after training, and feedforward the testing samples as a normal NN, which means that we just ignore the sequential dependencies in the testing phase. Finally, the AUC is 88.25\% and RIG is 18.95\%. Compared to Table~\ref{table:overall_performance}, we can observe a severe drop of the performance of RNN model, which shows that the sequential information is indeed embedded in the recurrent structure and significantly contribute to the prediction accuracy.
	
    \subsubsection{Performance with Long vs. Short History}
	
    In this part, we conduct experiments to check the model performance with different length of history. We first collect all available user sequences whose length is larger than a threshold $T$. In these sequences, the first $T$ samples in each sequences are fed into model to serve as the ``accumulation period''. Then, we continue feeding and testing samples for the rest part of each sequence, and calculate the AUC and RIG on all those rest parts. In such setting, the user sequences which are selected with a larger threshold $T$ have longer history to feed as ``accumulation period''. By doing so, we aim to verify whether our RNN model can maintain more robust sequential information in longer sequences.
   	
\begin{figure}
		\centering
		\includegraphics[width=.46\textwidth]{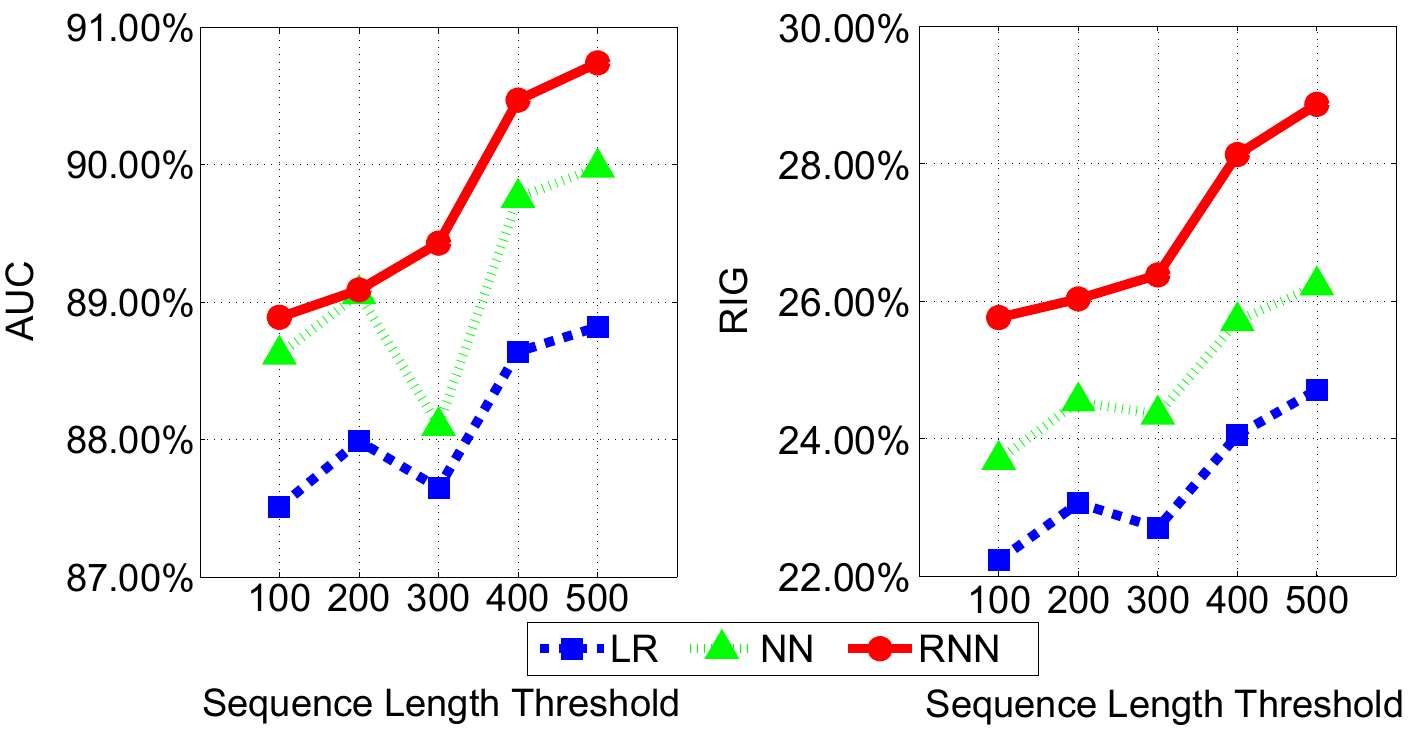}
		\caption{Performance with different history length $T$.}
		\label{fig:auc_rig_sequences}
    \end{figure}

    Figure~\ref{fig:auc_rig_sequences} shows the results. Just as expected, it turns out that our RNN model performs the best in all settings. Moreover, when the ``accumulation period'' gets longer, our RNN model tends to achieve even more relative gain compared to baseline models. This result validates the capability of RNN to capture and accumulate sequential information, especially for long sequences, and help further improve the accuracy of click prediction.
	
    \subsubsection{Effect of RNN Unfolding Step}

    As described in the framework section, the unfolding structure plays an import role in RNN training process with BPTT algorithm. Since unfolding can directly incorporate several previous samples, and the depth of such explicit sequence modeling is determined by the steps of unfolding, it is necessary to delve into the effect of various RNN unfolding steps. This further analysis can help us better understand the property of the RNN model.

    According to our experimental results, the prediction accuracy surges along with the increasing unfolding steps at the beginning. The best AUC and RIG can be achieved when unfolding 3 steps, after which the performance drops. By checking the error terms during the process of BPTT, we discover that the backpropagated error vanishes after 3 steps of unfolding, which explains why larger unfolding step is detrimental.

    With these observations, intuitively, our RNN based framework can model sequential dependency in two ways: short-span dependency by explicitly learning from current input and several of its leading inputs through unfolding; long-span dependency by implicitly learning from all previous input, accumulated or embedded in the weights of the recurrent part. Meanwhile, in sponsored search, user's behavior is also affected by both the very recent events as explicit factor and the long-run history as implicit (background) factor. This reveals the intrinsic reason why RNN works so well for the sequential click prediction.

\section{Conclusion and Future Work}\label{sec:conclusion}

    In this paper, we propose a novel framework for click prediction based on Recurrent Neural Networks. Different from traditional click prediction models, our method leverages the temporal dependency in user's behavior sequence through the recurrent structure. A series of experiments show that our method outperforms state-of-the-art click prediction models in various settings. In this future, we will continue this direction in several aspects: 1) The sequence is currently built on user level. We will study different kinds of sequence building methods, e.g. by (user, ad) pair, (user, query) pair, advertiser, or even merge all users on the level of whole system. 2) We are going to deduce the meaning of dependency learnt by RNN via deep understanding of RNN structure. This may help up better utilize the property of the recurrent part. 3) Recently, some research work~\cite{hermans2013training} has been done on Deep Recurrent Neural Networks (DRNN) and shows good results. We plan to study whether ``deep'' structure can also help in click prediction, together with the ``recurrent'' structure.

\clearpage

\bibliographystyle{aaai}
{
    \bibliography{References}
}

\end{document}